\documentclass[aps,prb,showpacs,twocolumn,floats,epsfig]{revtex4}
\usepackage{amssymb}
\usepackage{amsbsy}
\usepackage{amsmath}
\usepackage{epsfig}
\newcommand\beq{\begin{equation}}
\newcommand\eeq{\end{equation}}
\newcommand\bea{\begin{eqnarray}}
\newcommand\eea{\end{eqnarray}}
\newcommand\non{\nonumber}

\begin{document}

\title{Spin polarized STM spectra of Dirac Fermions on the surface of
a topological insulator}

\author{K. Saha$^{(1)}$, Sourin Das$^{(2)}$, K. Sengupta$^{(1)}$ and D.
Sen$^{(3)}$}
\affiliation{$^{(1)}$ Theoretical Physics Department, Indian Association
for the Cultivation of Science, Jadavpur, Kolkata 700 032, India \\
$^{(2)}$ Department of Physics and Astrophysics, University of Delhi,
Delhi 110 007, India \\
$^{(3)}$ Center for High Energy Physics, Indian Institute of Science,
Bangalore 560 012, India}

\date{\today}

\begin{abstract}

We provide a theory for the tunneling conductance $G(V)$ of Dirac
Fermions on the surface of a topological insulator as measured by a
spin-polarized scanning tunneling microscope tip for low bias
voltages $V$. We show that $G(V)$ exhibits an unconventional
dependence on the direction of magnetization of the tip and can be
used to measure the magnitude of the local out-of-plane spin
orientation of the Dirac Fermions on the surface. We also
demonstrate that if the in-plane rotational symmetry on the surface
of the topological insulator is broken by an external field, then
$G(V)$ acquires a dependence on the azimuthal angle of the
magnetization of the tip. We explain the role of the Dirac Fermions
in this unconventional behavior and suggest experiments to test our
theory.
\end{abstract}


\maketitle

\section{Introduction}

Topological insulators in both two and three dimensions (2D and 3D)
have attracted a lot theoretical and experimental attention in
recent years \cite{zhang1,hassan1,kane1,kane2,moore1,roy1,kon1}. It
has been shown in Refs. \onlinecite{kane2,moore1,roy1} that such 3D
insulators can be completely characterized by four integers $\nu_0$
and $\nu_{1,2,3}$. The former specifies the class of topological
insulators to be strong ($\nu_0=1$) or weak ($\nu_0=0$) while the
latter integers characterize the time-reversal invariant momenta of
the system given by $\vec M = (\nu_1 \vec b_1, \nu_2 \vec b_2, \nu_3
\vec b_3)/2$, where $\vec b_{1,2,3}$ are the reciprocal lattice
vectors. The latter integers $\nu_{1,2,3}$ are basis dependent while
the former, $\nu_0$, is a basis-independent invariant for a given
material. The weak topological insulators (WTI) are adiabatically
connected to Anderson insulators whereas the strong topological
insulators (STI) are not; consequently the topological features of
STI are robust against the presence of time-reversal invariant
perturbations such as disorder or lattice imperfections. STI exhibit
a host of novel phenomenon such as induction of magnetic monopoles
in the presence of an electric charge near its surface
\cite{zhangsc}, presence of topologically protected Fermion modes
inside dislocations \cite{ran1}, and the possibility of realization
of magnetic control over electrical conduction in its junction
\cite{mondal1}. It has been theoretically predicted
\cite{kane2,zhang1} and later experimentally verified \cite{hassan1}
that the surface of a STI has an odd number of Dirac cones whose
positions are determined by projection of $\vec M$ to the surface
Brillouin zone. The position and number of these cones depend on
both the nature of the surface concerned and the integers
$\nu_{1,2,3}$. For several compounds specific surfaces can be found
which will host a single Dirac cone near the $\Gamma$ point of the
2D Brillouin zone \cite{hassan1,exp2}. Such a Dirac cone at the
surface of a topological insulator is described by the Dirac
Hamiltonian \cite{kane4} \bea H_0 = \int \frac{d^2 k}{(2 \pi)^2}
\psi^{\dagger}(\vec k) \left[\hbar v_F \left({\vec \sigma} \times
{\vec k}\right)\cdot {\hat z} - \mu I\right] \psi(\vec k),
\label{dirham1} \eea where $\vec \sigma (I)$ denote the Pauli
(identity) matrices in spin space, $\psi(\vec k)=
(\psi_{\uparrow}(\vec k), \psi_{\downarrow}(\vec k))^T$ is the
electron annihilation operator, ${\vec k}=(k_x,k_y)$ is the 2D wave
vector, ${\hat z}$ denote the unit vector normal to the topological
insulator surface, and $\mu$ is the chemical potential. The
properties of these surface Dirac electrons has been studied in
detail in recent years. In particular, these Fermions are expected
to exhibit spin-momentum locking which predicts their spin to be
along the surface. This phenomenon has been extensively studied by
recent spin-resolved ARPES studies \cite{exp2,spar1}. However, in
many cases, such studies lead to a finite, albeit small, probability
of the measured spin polarization to point along the $\hat z$
direction \cite{spar2}. It is unclear at the moment whether this
contribution comes due to possible additional higher order terms in
the Dirac Hamiltonian of the surface electrons \cite{fu1} which can
spoil the spin-momentum locking property, or due to the presence of
additional scattering potentials induced by point-like or step-like
defects. Such spin-momentum locking for the surface electrons has
also been studied indirectly through scanning tunneling microscope
(STM) studies near step edges \cite{hari1}. However, the surface of
these insulators has not been studied using spin-polarized STM
\cite{comment1}.

\begin{figure}
\rotatebox{0}{
\includegraphics*[width=0.95 \linewidth]{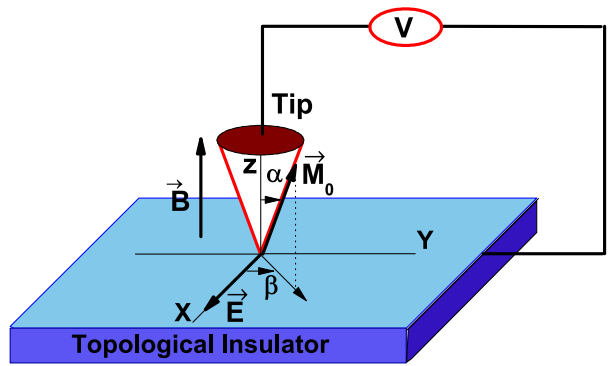}}
\caption{(Color online) Schematic representation of the proposed experimental
setup. The polar ($\alpha$) and the azimuthal ($\beta$) angles of the tip
magnetization $\vec M_0$ are shown. The applied electric and magnetic
field are along $\hat z$ and $\hat x$ respectively.} \label{fig1} \end{figure}

In this work we develop a theory for the tunneling current of the
Dirac Fermions on the surface of a topological insulator as measured
by a spin-polarized STM tip. We derive explicit expressions for the
tunneling current $I(V)$ and the tunneling conductance
$G(V)=dI(V)/dV$ and show that these quantities exhibit an
unconventional dependence on the direction of the tip magnetization.
Such an unconventional nature of the tunneling current and
conductance originates from the fact that Fermions on the surfaces
of topological insulators, in contrast to conventional electrons
obeying the Schr\"odinger equation, exhibit spin-momentum locking
which does not allow a free choice of their spin quantization axis
(which is fixed along the $\hat z$ direction as evident from Eq.\
(\ref{dirham1})). We demonstrate that the analysis of $G(V)$
provides a direct method of measurement of the local out-of-plane
spin component of the Dirac Fermions on the surface. Further, we
show that if the in-plane rotational symmetry on the surface of the
topological insulator is broken by an external field, then $G(V)$
acquires a dependence on the azimuthal angle of the tip
magnetization. We substantiate our theory and demonstrate these
unconventional features by a computation of $G(V)$ for Dirac
Fermions in the presence of a crossed electric (in-plane along $\hat
x$) and magnetic field (out of plane along $\hat z$) as shown in
Fig.\ (\ref{fig1}). We suggest experiments which can test our
theory.

The organization of the rest of the paper is as follows. In Sec.\
\ref{form1}, we obtain the general expressions for $I(V)$ and $G(V)$
as measured by a spin-polarized STM tip at low bias voltages $V$.
This is followed by Sec.\ \ref{res} where we provide an explicit
calculation of the tunneling conductance $G(V)$ for Dirac Fermions
on the surface of a topological insulator in the presence of a
crossed in-plane electric and out-of-plane magnetic field. We
discuss possible experiments and conclude in Sec.\ \ref{conc1}.

\section{Tunneling current due to a spin-polarized STM}
\label{form1}

The experimental situation that we intend to describe is
schematically shown in Fig.\ (\ref{fig1}). The STM tip is placed
atop the surface of the topological insulator and has a
magnetization $\vec M_0$ as shown in Fig.\ (\ref{fig1}). To
understand the unconventional behavior of the tunneling current in
such a system, let us briefly review the current obtained by such a
spin-polarized tip when placed atop the surface of a conventional
magnetic material with a magnetization $\vec m_0(\bf r)$. In this
case, the STM current is given by the well known expression
\cite{magstm1,chen1} \bea I(V) &=& \frac{2 \pi e}{\hbar} \sum_{\mu
\nu} \delta \left(E_{\mu}^t-E_{\nu}^s-eV\right)
\left[f\left(E_{\mu}^t\right)-
f\left(E_{\nu}^s\right)\right] \non \\
&& ~~~~~~~~~~~\times |M_{\mu \nu}|^2 , \label{stm1} \eea where $f$
is the Fermi distribution function, $E_{\mu}^t$ and $E_{\nu}^s$
denote the energy levels of the tip and the surface states
respectively, and $M_{\mu \nu}$ denotes the matrix element for
overlap between the tip and the STM wave functions which is
determined by the Bardeen tunneling formula \cite{bardeen1} \bea
M_{\mu \nu} &=& \frac{\hbar^2}{2m} \int_s ds
\left[\psi_{\mu}^{t\,\ast}({\bf r},z)\partial_z \psi_{\nu}^s ({\bf
r},z) \right. \nonumber\\
&& \left. - \psi_{\nu}^s ({\bf r},z)
\partial_z \psi_{\mu}^{t\,\ast}({\bf r},z)\right],
\label{stm2} \eea where $\bf r$ denotes only the $(x,y)$
coordinates. Here $\psi_{\mu}^t$ and $\psi_{\nu}^s$ denote the wave
functions of the electrons in the tip and on the sample surface
respectively, $z$ is taken to be the direction perpendicular to the
surface, and the integral is to be performed over a plane which lies
between the STM tip and the sample surface and is parallel to the
sample surface.

For a conventional magnetic sample, one can always choose the local
spin quantization axis to be along the direction of magnetization of
the tip. Consequently the wave function of the tip electrons can be
written as \bea \psi_{\mu}^t({\bf r};z) &=& (1,0) \chi_{\mu}({\bf
r};z), \label{tipwave1} \eea where the precise form of
$\chi_{\mu}({\bf r})$ is difficult to obtain theoretically and it is
usually modeled to have the same orbital symmetry of the electron
for the atom at the STM tip. For the rest of this work, we shall
assume an $s$-wave symmetric wave function for the tip electrons;
all the results obtained can be extended to other orbital symmetries
in a straightforward manner using the methods of Ref.\
\onlinecite{chen1}. Note that such a choice of the spin quantization
axis of the tip is possible since both the surface and the tip
electrons obey Schr\"odinger equations which, in contrast to the
Dirac equation, do not specify a choice of the spin quantization
axis. For the surface electrons, the wave function can be written as
\bea \psi_{\nu}^s({\bf r},z) &=&(\cos(\theta_{\nu}({\bf
r})/2),\sin(\theta_{\nu}({\bf
r})/2)e^{i\phi_{\nu}({\bf r})})\xi_{\nu}({\bf r};z) \non \\
\label{surfacewave1} \eea where the angles $\theta_{\nu}({\bf r},z)$
and $\phi_{\nu}({\bf r},z)$ are polar and azimuthal angles with
respect to the tip magnetization, and $\xi_{\mu}({\bf r})$ denotes
the spatial part of the surface electron wave function whose
specific form depends on the system \cite{ter1,chen1}. Using Eqs.\
(\ref{stm2}), (\ref{tipwave1}) and (\ref{surfacewave1}), a
straightforward analysis following Ref.\ \onlinecite{ter1} yields
\bea M_{\mu \nu}= c_0 \cos(\theta_{\nu}({\bf r_t})/2) \xi_{\nu}({\bf
r_t}), \label{melement3} \eea where ${\bf r_t} \equiv({\bf
r}_t,z_t)$ denotes the position of the tip, and the constant $c_{0}$
depends on the details of the tip wave function whose expression for
a spherically symmetric s-wave tip is given by
\cite{magstm1,bardeen1,ter1} \bea c_{0} = \frac{2\pi N
\hbar^2}{\kappa m}, \label{consteq} \eea where $N$ is the
normalization constant, $\kappa$ is the decay length of the tip
wavefunction, and $m$ is the mass of the tip electrons. Substituting
Eqs.\ (\ref{melement3}) in Eq.\ (\ref{stm1}) one obtains a more
familiar expression for the current \bea I(V) &=& I_0 |c_0|^2 \rho_t
\left(\rho_s(eV;{\bf r_t})+ m_s(eV,{\bf r_t})\right),
\label{magcurr1} \eea where $V$ is the applied voltage, $I_0 = 2\pi
e/\hbar$, $\rho_t$ is the density of states for the tip electron
which is assumed to be constant within the range of experimental
applied voltages, and the integrated local density of states
$\rho_s\equiv \rho_s(eV;{\bf r_t})$ and magnetization $m_s \equiv
m_s(eV;{\bf r_t})$ are given by
\bea \rho_s &=& \int d\omega \left[f(\omega)-f(\omega+eV)\right] \non \\
&& \times \sum_{\nu} \delta(E_{\nu} -\omega) |\xi_{\nu}({\bf r_t})|^2 \non \\
m_s &=& \int d\omega \left[f(\omega)-f(\omega+eV)\right] \non \\
&& \times \sum_{\nu} \delta(E_{\nu} -\omega) |\xi_{\nu}({\bf
r_t})|^2 \cos(\theta_{\nu} ({\bf r_t})). \label{idos1} \eea Note
that if we restrict ourselves to low bias voltages and low
temperatures for which the contribution to $m_s$ comes from a narrow
range of $\nu$ for which the relative angle of magnetization between
the sample and the STM tip is approximately a constant, we obtain
\bea I(V)=I_0 |c_0|^2 \rho_t \rho_s(eV;{\bf r}_t) \left[ 1+
\cos(\theta({\bf r_t})) \right] \label{magcurrent2} \eea which is
the well-known expression for the current for a magnetic material as
measured by a fully polarized STM tip.

The central result of our work, which we now derive, is that Eq.\
(\ref{magcurrent2}) needs to be significantly modified for the
calculation of $I(V)$ for a topological insulator. For a topological
insulator, the electrons on the surface are described by the Dirac
Hamiltonian given by Eq.\ (\ref{dirham1}) with some possible
additional warping terms, impurity potentials, and/or external
electric and magnetic fields. In all such cases, the spin
quantization axis of the electrons is fixed to be along $\hat z$. In
this situation, for the system of polarized STM tip near the
surface, once we choose to describe the Dirac Fermions by Eq.\
(\ref{dirham1}) (with possible additional terms), it is no longer
possible to fix the spin quantization axis of the system along the
magnetization of the STM tip. Thus the tip wave function has to be
represented as \bea \psi_{\mu}^t({\bf r};z) &=&
\left(\cos(\alpha_{\mu}({\bf r},z)/2),\sin(\alpha_{\mu}({\bf
r},z)/2)e^{i \beta_{\mu}({\bf
r},z)}\right) \non \\
&& \times \chi_{\mu}({\bf r};z), \label{tipwave2} \eea where
$\alpha_{\mu}({\bf r},z)$ and $\beta_{\mu}({\bf r},z)$ denotes the
polar (as measured with respect to $\hat z$) and azimuthal angles of
the magnetization (as measured with respect to an in-plane
rotational symmetry breaking field along $\hat x$ which will be
specified later) as shown in Fig.\ (\ref{fig1}). Note that in the
absence of an in-plane symmetry breaking field, as shown in Fig.\
(\ref{fig2}), the azimuthal angle can be chosen to have any
arbitrary value. The wave function of the Dirac electrons on the
surface of a topological insulator in real space can be written as
\bea \psi_{\nu}^{\rm TI}({\bf r};z) &=& \left(\psi_{\uparrow
\nu}^{\rm
TI}({\bf r};z), \psi_{\downarrow \nu}^{\rm TI}({\bf r};z) \right) \non \\
&=& \left(1, \kappa_{\nu}({\bf r},z)e^{-i \eta_{\nu}({\bf
r},z)}\right) \psi_{\uparrow \nu}^{\rm TI}({\bf r};z)
\label{tiwave1} \eea where $\psi_{\uparrow \nu}^{\rm TI}({\bf r};z)$
and $\psi_{\downarrow \nu}^{\rm TI}({\bf r};z)$ represent the
spin-up and spin-down component of the Dirac electrons and we have
used $\psi_{\downarrow \nu}^{\rm TI}({\bf r};z)/ \psi_{\uparrow
\nu}^{\rm TI}({\bf r};z) =\kappa_{\nu}({\bf r},z)\exp[-i
\eta_{\nu}({\bf r},z)]$. We will provide explicit expressions for
these wave functions for a specific case in the next section.

Substituting Eqs.\ (\ref{tipwave2}) and (\ref{tiwave1}) in Eq.\
(\ref{stm2}), one then obtain the matrix element $M_{\mu \nu}$
entering the expression of the current $I(V)$ in Eq.\ (\ref{stm1})
to be \bea M_{\mu \nu} &=& c_{0} \psi_{\uparrow \mu}^{\rm TI}({\bf
r};z) \Big[ \cos(\alpha_{\mu}({\bf r},z)/2) + \kappa_{\nu}({\bf r},z) \non \\
&& \times \sin(\alpha_{\mu}({\bf r},z)/2)e^{i [\beta_{\mu}({\bf
r},z)-\eta_{\nu}({\bf r},z)]} \Big] \label{melement2} \eea which
yields, assuming constant density of states for the tip and energy
independence of the tip magnetization angles $\alpha$ and $\beta$,
\bea I(V)&=& I_0 |c_0|^2 \rho_t \Big[\rho_{d}(eV;{\bf r_t}) + \rho_z
(eV;{\bf r_t}) \cos(\alpha) \non \\
&& + \rho_{m}(eV, \beta;{\bf r_t}) \sin(\alpha) \Big] \non \\
\rho_{d}(eV;{\bf r}_t) &=& \frac{1}{2}\int d\omega
\left[f(\omega)-f(\omega+eV)\right] \non \\
&& \times \sum_{\nu} \delta(E_{\nu} -\omega) |\psi_{\uparrow
\nu}^{\rm TI}({\bf r_t})|^2[1+\kappa_{\nu}^2({\bf r_t})] \non \\
\rho_{z}(eV;{\bf r}_t) &=& \frac{1}{2} \int d\omega
\left[f(\omega)-f(\omega+eV)\right] \non \\
&& \times \sum_{\nu} \delta(E_{\nu} -\omega) |\psi_{\uparrow
\nu}^{\rm TI}({\bf r_t})|^2[1-\kappa_{\nu}^2({\bf r_t})] \non \\
\rho_{m}(eV,\beta;{\bf r}_t) &=& \int d\omega
\left[f(\omega)-f(\omega+eV)\right] \sum_{\nu} \delta(E_{\nu} -\omega) \non \\
&& \times |\psi_{\uparrow \nu}^{{\rm TI}}|^2 \kappa_{\nu}({\bf r_t})
\cos[\beta-\eta_{\nu}({\bf r_t})] \label{stmfinal} \eea Using Eq.\
\ref{stmfinal}, the tunneling conductance $G(V)$ at $T=0$ can be
obtained from Eq.\ (\ref{stmfinal}) as \bea G&=& G_0 |c_0|^2 \rho_t
\Big[\rho'_{d}(eV;{\bf r_t}) + \rho'_z (eV;{\bf
r_t}) \cos(\alpha) \non \\
&& + \rho'_{m}(eV, \beta;{\bf r_t}) \sin(\alpha) \Big] \non \\
\rho'_{d}(eV;{\bf r}_t) &=& \frac{1}{2} \sum_{\nu} \delta(E_{\nu}
-eV) |\psi_{\uparrow \nu}^{\rm TI}({\bf
r_t})|^2[1+\kappa_{\nu}^2({\bf r_t})] \non \\
\rho'_{z}(eV;{\bf r}_t) &=& \frac{1}{2}\sum_{\nu} \delta(E_{\nu}
-eV) |\psi_{\uparrow \nu}^{\rm TI}({\bf
r_t})|^2[1-\kappa_{\nu}^2({\bf r_t})] \non \\
\rho'_{m}(eV,\beta;{\bf r}_t) &=& \sum_{\nu} \delta(E_{\nu} -eV)
|\psi_{\uparrow \nu}^{{\rm TI}}|^2 \kappa_{\nu}({\bf r_t}) \non \\
&& \times \cos[\beta-\eta_{\nu}({\bf r_t})], \label{condeq} \eea
where $G_0 = 2\pi e^2/\hbar$.

Eqs.\ (\ref{stmfinal}) and (\ref{condeq}) represent the central
result of this work. First, we note that the tunneling conductance
displays an unconventional dependence on both the polar and
azimuthal angles of the tip magnetization. In particular, $G(V)$ may
exhibit a dependence on the azimuthal angle $\beta$ of the
magnetization of the tip which never occurs for STM spectra of
conventional magnetic samples. Second, we note that we do not expect
this unconventional dependence to show up when the tip is placed on
top of a pristine topological insulator surface without any defects
and/or in the absence of external fields. In this case, since the
spins of the Dirac electrons point along the plane due to
spin-momentum locking, one expects $\kappa_{\nu}({\bf r_t})=1$ which
leads to $\rho_z=0$. Further, since the sample is completely
isotropic, the sum over all the surface states with a given energy
$E_{\nu}$ in Eq.\ (\ref{stmfinal}) will make the current vanish.
This can be seen from the fact that such a sum must be independent
of the orientation of the $\hat x$ and $\hat y$ axis on the surface
of the TI; therefore shifting $\eta_\nu \to \eta_\nu + c$ should not
change any of the results. Hence, the quantity $\rho_m$ in Eq.\
(\ref{stmfinal}) must be independent of the choice of $\beta$ and
must therefore vanish. This leads to $I(V) = I_0 |c_0|^2 \rho_t
\rho_d(eV;{\bf r_t})$ which is independent of both $\alpha$ and
$\beta$. Finally, we would like to point out that the presence of a
non-zero component of $S_z$ due to warping \cite{fu1} or other
effects such as presence of defects \cite{hari1} renders
$\kappa_{\nu}({\bf r_t}) \ne 1$ and hence leads to a $\cos(\alpha)$
dependence of $I(V)$ and $G(V)$. Thus measuring $G(V)$ with a
spin-polarized STM provides a direct way of measuring the real-space
out-of-plane spin polarization of the Dirac electrons on the surface
of a topological insulator. We shall discuss this point in more
details in Sec.\ \ref{conc1}.

In the next section, we compute $G(V)$ for the Dirac electrons in
the presence of a crossed electric (along $\hat x$) and magnetic
field (along $\hat z$) and show that $G(V)$ displays all the
unconventional features discussed above.

\section{Crossed electric and magnetic field}
\label{res}

In this section, we compute $G(V)$ for the geometry shown in Fig.\
(\ref{fig1}), {\it i.e.}, in the presence of a constant magnetic
field $B$ along $\hat z$ and an electric field $E$ along $\hat x$.
Throughout this section, we shall assume that the applied magnetic
field is weak enough so as not to orient the magnetization of the
tip along $B$. For a stronger magnetic field, one needs to apply the
magnetic field along the direction of the magnetization of the tip.
Note that for the topological insulator, this just amounts to
changing $B \to B_z=B \cos(\alpha)$ since the in-plane component of
the field appears to be a gauge shift for the Dirac electrons and
does not have any influence on their properties \cite{mondal1}. So
all of the analysis of the present section will hold in this case,
with $B \to B \cos(\alpha)$.

We first look for the solution of the wave function $\psi_{\nu
\uparrow, \downarrow}^{\rm TI}({\bf r},z)= \psi_{\nu \uparrow,
\downarrow}^{\rm TI}({\bf r}) \exp(-\kappa_z z)$, where $\kappa_z$
is the decay length of the wave function in real space which depends
on the work function of the topological insulator surface. For
estimating $G(V)$, we need the value of the wave function at the
position of the tip, ${\it i.e.}$, at $z=z_t$. Thus it is possible
to absorb $\exp(-\kappa_z z_t)$ in the definition of $G_0$ and one
does not need to know the precise value of $\kappa_z$ to estimate
$G$. The solution for $\psi_{\mu \uparrow, \downarrow}^{\rm TI}({\bf
r})$ and the corresponding energies $E_{\nu}$ for the present
configuration is well-known \cite{mondal1,shankar1}. We will assume
that the electric field is smaller than the magnetic field ($E < v_F
B$); then one can use a Lorentz transformation to go to a frame
where the electric field is zero, solve the Landau level problem in
that frame, and then Lorentz transform back to the original frame.
The energy eigenstates are labeled by the Landau level index $n$ and
the transverse momenta $k_y$ and can be expressed in terms of
$\lambda={\rm arccosh}[(1-E^2/v_F^2 B^2)^{-1/2}]$ for $n\ne 0$ by
\bea \psi_{n,k_y,\uparrow}^{\rm TI}({\bf r}) &=& \left[
\cosh(\lambda/2)
\chi_{n-1}(\xi) + i \sinh(\lambda/2) \chi_{n}(\xi) \right] e^{i k_y y}, \non \\
\psi_{n,k_y,\downarrow}^{\rm TI}({\bf r})&=& \left[ \cosh(\lambda/2)
\chi_{n}(\xi) - i \sinh(\lambda/2) \chi_{n-1}(\xi) \right] e^{ik_y y}, \non \\
E_{n,k_y} &=& {\rm Sgn}(n) E_0\cosh(\lambda)^{-3/2}\non \\
&& \times \sqrt{|n|+ c_z B\cosh(\lambda)} - \hbar k_y E/B,
\label{waven1} \eea where $E_0=\hbar v_F l_B^{-1}$,
$\chi_n=\exp(-\xi^2/2) H_n(\xi)/\sqrt{2^{|n|} |n|!}$, $H_n$ denotes
Hermite polynomials, $l_B=\sqrt{\hbar c/eB}$ is the magnetic length,
$c_z= g^2 \mu_B^2/(\hbar v_F^2 e) $, $g$ is the gyromagnetic ratio,
$\mu_B$ is the Bohr magneton, and $\xi$ is given by \bea \xi= (x
-k_y l_B^2)/\sqrt{\gamma l_B^2 } +\sqrt{2|n|} E/(v_F B).
\label{xieq} \eea {}From Eq.\ (\ref{waven1}), one obtains the
parameters $\kappa$ and $\eta$ to be \bea \kappa_{n,k_y}(x) &=&
\frac{\sqrt{\cosh^2(\lambda/2)\chi^2_{n}(\xi)+ \sinh^2(\lambda/2)
\chi^2_{n-1}(\xi)}}{\sqrt{\cosh^2(\lambda/2)\chi^2_{n-1}(\xi)+
\sinh^2(\lambda/2)
\chi^2_n(\xi)}}, \non \\
\eta_{n,k_y} (x) &=& \arctan \left[\tanh(\lambda/2)\chi_n(\xi)/
\chi_{n-1}(\xi)\right] \non \\
&& + \arctan \left[
\tanh(\lambda/2)\chi_{n-1}(\xi)/\chi_n(\xi)\right]. \label{kappaeta}
\eea The corresponding expressions for the $n=0$ Landau level are
given by
\bea E_{n,k_y}&=& -g \mu_B B/ \cosh(\lambda)- \hbar k_y E/B, \non \\
\psi_{n,k_y,\uparrow}^{\rm TI}({\bf r}) &=& e^{i k_y y} i \sinh(\lambda/2)
\chi_0(\xi), \non \\
\psi_{n,k_y,\downarrow}^{\rm TI}({\bf r})&=& e^{i k_y y}
\cosh(\lambda/2) \chi_0(\xi), \label{waven2} \eea
which leads to $\kappa_0=\coth(\lambda/2)$ and $\eta_0=-\pi/2$
independent of $k_y$. We note here that for typical topological
insulators $g \mu_B B/E_0 \sim 10^{-4}$ for $B \simeq 1T$ and so the
Zeeman term can be safely neglected in all numerical estimates of $G$.

\begin{figure}
\rotatebox{0}{
\includegraphics*[width=0.95 \linewidth]{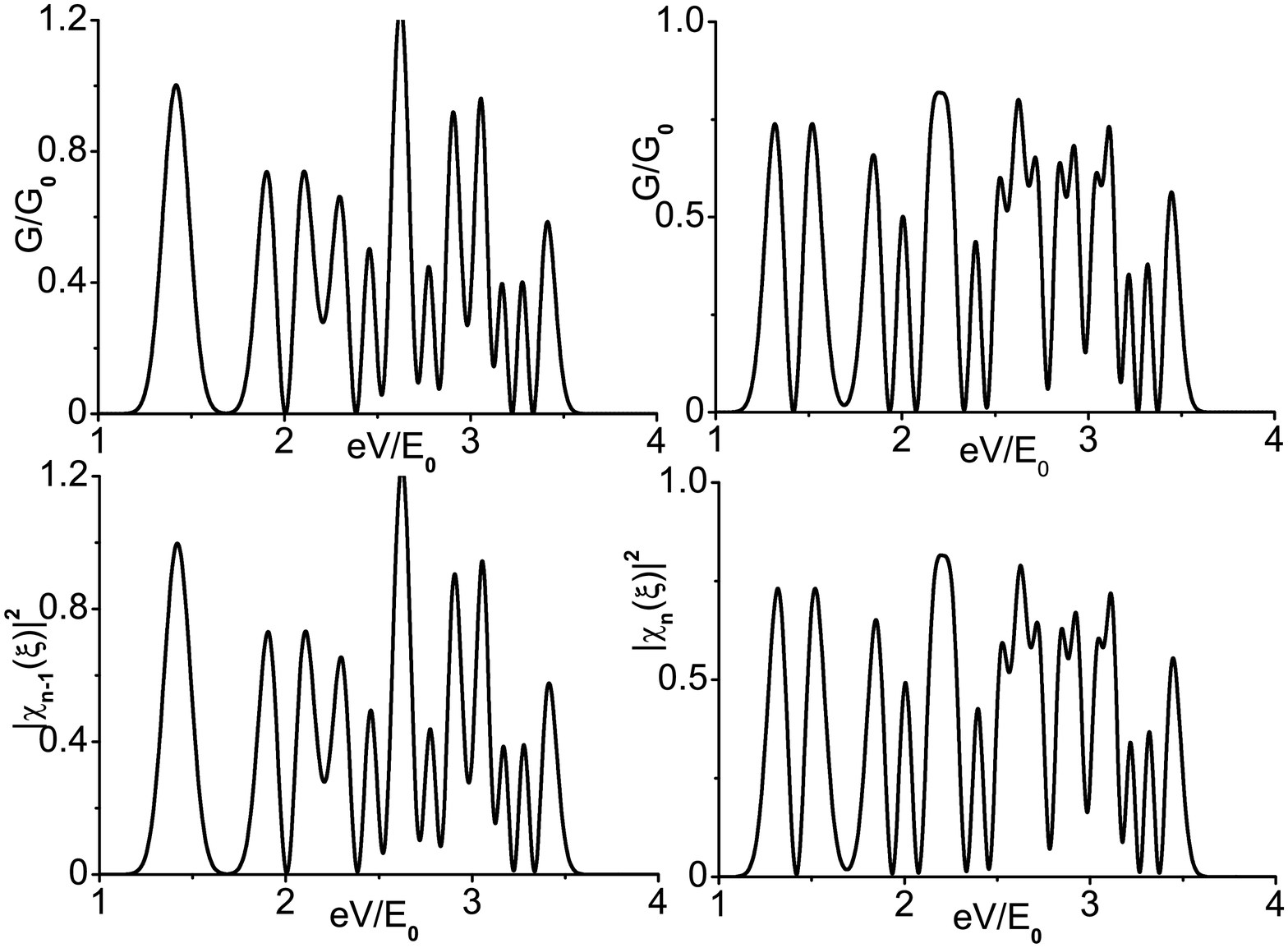}}
\caption{(Color online) Plot of $G/G_0$ as a function of the applied
voltage $eV/E_0$ for $\alpha=0$ (top left panel) and $\alpha=\pi$
(top right panel) for $E/v_F B=0.1$ and $|c_0|^2 \rho_t/E_0=1$. The
bottom panels show the variation of the wave functions
$|\chi_{n-1}(\xi)|^2$ (bottom left) and $|\chi_n(\xi)|^2$ (bottom
right) corresponding to the solution $(n,k_y)$ of $eV=E_n(k_y)$
which contributes maximally to $G$. See text for details.}
\label{fig2} \end{figure}

To obtain $G$, we substitute Eqs.\ (\ref{waven1}) and
(\ref{kappaeta}) in Eq.\ (\ref{condeq}). In what follows, we choose
the STM tip to be directly on top of the origin of coordinates so
that ${\bf r_t}=(0,0)$. We note that corresponding to any given
applied voltage $V$, the solution of $E_{n, k_y}=eV$ constitutes, in
principle, an infinite number of pairs $(n, k_y)$. However, the
Gaussian factor $\exp(-\xi^2/2)$ in the expression of
$\psi_{n,k_y,\uparrow}^{\rm TI}({\bf r})$ and the discrete values of
$n$ ensures that out of all these pairs, the one with the minimal
value of $|k_y|$ provides the most significant contribution to $G$.
This is particularly valid for small $E \ll v_F B$ which is also the
regime we are interested in. In this regime, the tunneling
conductance $G(V)$ as a function of $V$, as computed using Eqs.\
(\ref{waven1}), (\ref{waven2}) and (\ref{condeq}), is shown in the
top left (right) panel of Fig.\ (\ref{fig2}) for $E/v_F B=0.1$ and
$\alpha=0 ~(\pi)$. From Eq.\ (\ref{condeq}), we find that for both
$\alpha=0$ and $\alpha=\pi$, $\rho'_m=0$, so that one has \bea
G(V;\alpha=0) &=& G_0 |c_0|^2 \rho_t \sum_{\nu} \delta(E_{\nu} -eV)
|\psi_{\uparrow \nu}^{\rm TI}(0)|^2, \non \\
G(V;\alpha=\pi) &=& G_0 |c_0|^2 \rho_t \sum_{\nu} \delta(E_{\nu}
-eV) |\psi_{\downarrow \nu}^{\rm TI}(0)|^2. \non\\ \eea The nature
of $G(V)$ can now be understood from Eqs.\ (\ref{waven1}) and
(\ref{waven2}). From these equations, we find that as $V$ is changed
so that it deviates from the energy of a Landau level, the lowest
value of $k_y\equiv k_y^0$ which is a solution to $eV=E_{n,k_y}$
starts to increase. This leads, through a change of $\xi(k_y^0)$, to
a non-monotonic change in the wave functions $|\psi_{\uparrow
\nu}^{\rm TI}(0)|^2 $ (for $\alpha=0$) and $|\psi_{\downarrow
\nu}^{\rm TI}(0)|^2 $ (for $\alpha=\pi$) since these wave functions
involve a product of the Gaussian factor $\exp(-\xi^2/2)$ and the
Hermite polynomials $H_{n-1}(\xi)$ and $H_n(\xi)$. Note that for
$E/v_FB=0.1$, the predominant contribution to $|\psi_{\uparrow
\nu}^{\rm TI}(0)|^2 [|\psi_{\downarrow \nu}^{\rm TI}(0)|^2]$ comes
from the coefficients of the $\cosh(\lambda/2)$ terms and thus
involve $H_{n-1}(\xi) ~[H_n(\xi)]$ for $\alpha=0 ~[\pi]$. In fact,
the plots of $\chi_{n-1}[\xi(k_y^0)] ~(\chi_n[\xi(k_y^0)])$ as a
function of $V$ in the lower left(right) panel of Fig.\ (\ref{fig2})
(for $\alpha=0 ~(\pi)$) show that the non-monotonic behavior of
$G(V)$ can be understood in terms of the variation of these wave
functions with the change of $k_y^0$ as a function of $V$. We note
that the spectral features in $G(V)$ are expected to change
drastically with the direction of the tip magnetization as is
evident from comparing the left and right panels of Fig.\
(\ref{fig2}).

\begin{figure}
\rotatebox{0}{
\includegraphics*[width=0.95 \linewidth]{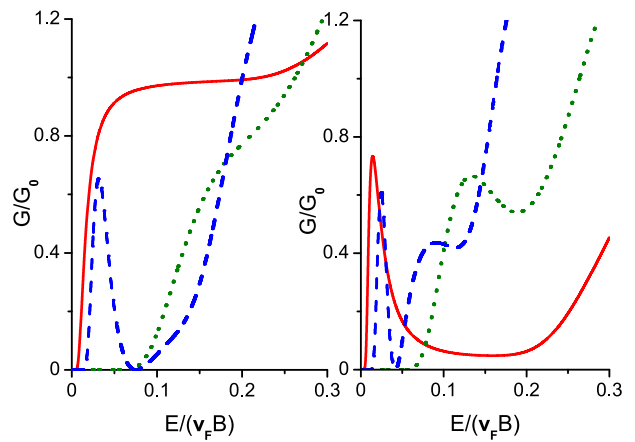}}
\caption{(Color online) Plot of $G/G_0$ as a function of the applied
electric field $E/v_F B$ for several applied voltages $eV/E_0=1.4$
(red solid line), $1.8$ (green dotted line), and $2.4$ (blue dashed
line) for which the maximally contributing $(n,k_y)$ pair
corresponds to $n=1,~2 ~{\rm and}~ 3$. The left (right) panel
corresponds to $\alpha=0 ~(\pi)$. We have taken $|c_0|^2
\rho_t/E_0=1$ in all the plots. See text for details.} \label{fig3}
\end{figure}

Next, we fix the applied voltage $eV/ E_0$ and study the behavior of
$G(V)$ as a function of the electric field $E/v_F B$ for several
values of $eV/E_0$ for $\alpha=0$ (left panel) and $\pi$ (right
panel) as shown in Fig.\ (\ref{fig3}). The nature of these plots can
again be understood from the variation of the wave functions
$\psi_{\uparrow \nu}^{\rm TI}(0)$ and $\psi_{\downarrow \nu}^{\rm
TI}(0)$. For example, let us consider the plot corresponding to
$eV/E_0=1.4$ (red solid lines in Fig.\ (\ref{fig3})) for which the
maximally contributing $(n,k_y)$ pair corresponds to $n=1$. As one
increases $E/v_F B$, we find from Eq.\ (\ref{xieq}) that the value
of $\xi$ decreases since the decrease in $k_y$ overcompensates for
the increase in $E$ leading to a net decrease in $\xi$ in the regime
where $E/v_F B \ll 1$. Consequently, for $\alpha=0$, where $G$
depends on $|\psi_{\uparrow \nu}^{\rm TI}(0)|^2 \sim |\chi_0(\xi)|^2
\sim \exp(-\xi^2)$, we find an increase in $G(V)$. For $\alpha=\pi$,
where $G$ depends on $|\psi_{\downarrow \nu}^{\rm TI}(0)|^2 \sim
|\chi_1(\xi)|^2 \sim |\exp(-\xi^2/2) H_1(\xi)|^2$, a similar
increase in $\xi$ leads to a non-monotonic behavior of $G$. The
plots of $G$ for other values of $eV/E_0$ shown in Fig.\
(\ref{fig3}) can be understood in a similar manner.

\begin{figure}
\rotatebox{0}{
\includegraphics*[width=0.95 \linewidth]{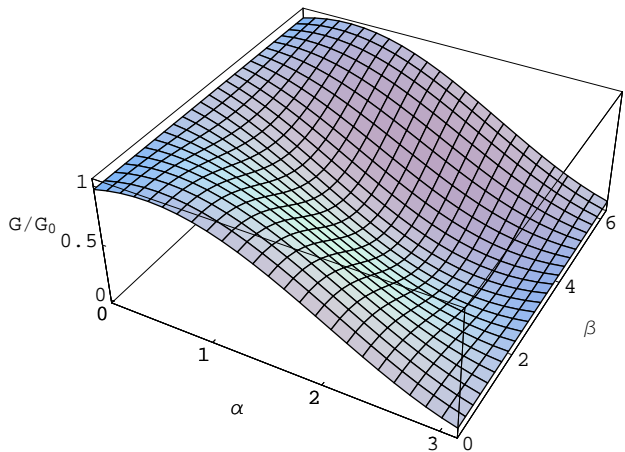}}
\caption{(Color online) Plot of $G/G_0$ as a function of the angles
$\alpha$ and $\beta$ for $E/v_F B=0.1$, $|c_0|^2 \rho_t/E_0=1$, and
$eV/E_0=1.4$.} \label{fig4}
\end{figure}

Finally, we consider the variation of the tunneling conductance as a
function of the polar ($\alpha$) and azimuthal ($\beta$) angles
corresponding to the direction of the tip magnetization for an
applied voltage $eV/E_0=1.4$ and electric field $E/v_F B=0.1$ as
shown in Fig.\ (\ref{fig4}). We note that $G(V)$ shows maximal
variation with $\beta$ for $\alpha=\pi/2$ as expected from the
expression of $\rho'_m$ in Eq.\ (\ref{condeq}). Such a variation is
highlighted in Fig.\ (\ref{fig5}) where the plot of $G'/G_0$ (where
$G'= \partial G/\partial \beta$) is shown as a function of $\beta$
for $\alpha=\pi/2$. We note that since $G'(\beta=0, \alpha=\pi/2)
\sim \sin(\eta({\bf r_t};V))$, the intercept of this plot is a
measure of the local relative phase between the spin-up and
spin-down components of the surface Dirac Fermions. Such a variation
of the tunneling conductance with the azimuthal angle of the tip
magnetization is a novel feature of the topological insulators which
originates from fixing the spin quantization axis of the surface
Dirac electrons along $\hat z$, and it has no analog in conventional
magnets. We note that the measurement of such a variation in an
experiment does not require a change in the direction of the tip
magnetization. It can simply be achieved by changing the direction
of the electric field $E$ which breaks rotational symmetry since
$\beta$ is always defined with respect to the direction of $E$.

\section{Experiments}
\label{conc1}

\begin{figure}
\rotatebox{0}{
\includegraphics*[width=0.95 \linewidth]{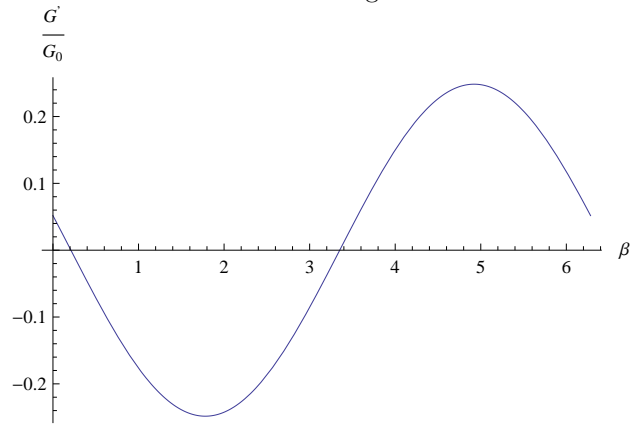}}
\caption{(Color online) Plot of $G'(V)/G_0$ (where $G'=\partial
G/\partial \beta$) as a function of $\beta$ for $E/v_F B=0.1$,
$|c_0|^2 \rho_t/E_0=1$, $\alpha=\pi/2$, and $eV/E_0=1.4$.}
\label{fig5}
\end{figure}

The experimental test of our work involves measurement of the STM
spectra of a topological insulator using a magnetic tip. We note
that such measurements can be used to map the spatial out-of-plane
spin-orientation profile of the surface Dirac electrons. In
particular, we suggest measurement of the spectra with the STM tip
magnetization pointing along $\hat z$ ($\alpha=0$) and $-\hat z$
($\alpha=\pi$). The local $\hat z$ component of the spin of the
Dirac electron can then be obtained from \bea S_z (V) ~\sim~
\frac{G(V;\alpha=0)-G(V;\alpha=\pi)}{G(V;\alpha=0)+
G(V;\alpha=\pi)}. \eea We note that such a measurement would provide
direct information about the real-space out-of-plane spin profile of
the surface Dirac electrons near surface defects such as point
impurities and step edges. The other experiment we suggest involves
measurement of the STM spectra with a tip whose magnetization points
along the surface of the topological insulator ($\alpha=\pi/2$) in
the presence of a crossed electric and magnetic field. Our central
prediction is that the tunneling conductance will display an
oscillatory behavior with the change of the azimuthal angle $\beta$
(which can be varied by changing the direction of the applied
electric field for a fixed tip magnetization) as can be seen from
plot of $\partial G/\partial \beta$ as a function of $\beta$ in
Fig.\ (\ref{fig5}). We also note that the value of $G'(\beta=0)$
provides direct information about the local relative phase of
between the spin-up and spin-down components of the electron
wavefunction on the topological insulator surface.

In conclusion, we have presented a general theory for the STM
spectra of Dirac electrons on the surface of a topological insulator
as measured by a magnetic STM tip and have shown that such a spectrum
has unconventional features not commonly seen in its counterpart in
conventional magnets. We have identified the reason for such an
unconventional spectrum to be the fixation of the spin quantization
axis of the surface Dirac electrons along the $\hat z$ direction. We have
also explicitly computed the STM spectra for the surface Dirac electrons
in the presence of a crossed electric and magnetic field which
demonstrates this unconventional behavior.

KS thanks H. Manoharan, G. Refael, N-C Yeh, and E. Zhao for several
related discussions and DST, India for support through grant no.
SR/S2/CMP-001/2009. DS thanks DST, India for financial support under
SR/S2/JCB-44/2010.

\end{document}